\documentclass{article}

\makeatletter
\renewcommand*{\@fnsymbol}[1]{\ifcase#1\or\dagger\or\@arabic{\numexpr#1-1\relax}\else\@arabic{\numexpr#1\relax}\fi}
\makeatother

\usepackage{arxiv}

\usepackage[utf8]{inputenc} 
\usepackage[T1]{fontenc}    
\usepackage{hyperref}       
\usepackage{url}            
\usepackage{booktabs}       
\usepackage{amsfonts}       
\usepackage{nicefrac}       
\usepackage{microtype}      
\usepackage{lipsum}
\usepackage{graphicx}
\usepackage{natbib}
\usepackage{amsmath}
\usepackage{cleveref}
\graphicspath{ {./images/} }
\usepackage[dvipsnames]{xcolor}

\definecolor{GoogleBlue}{HTML}{4285F4}
\definecolor{GoogleRed}{HTML}{EA4335}
\definecolor{GoogleYellow}{HTML}{FBBC05}
\definecolor{GoogleGreen}{HTML}{34A853}

\newcommand{\Google}{\textcolor{GoogleBlue}{G}\textcolor{GoogleRed}{o}\textcolor{GoogleYellow}{o}\textcolor{GoogleBlue}{g}\textcolor{GoogleGreen}{l}\textcolor{GoogleRed}{e} }

\title{Why all roads don’t lead to Rome:\\Representation geometry varies across the human visual cortical hierarchy}

\author{
 Arna Ghosh \thanks{$^{, \ddag}$ Equal Contribution\;\;\;}\; \thanks{Mila - Quebec AI Institute, Montr\'eal, QC, Canada\;\;\; $^2$ Computer Science, McGill University, Montr\'eal, QC, Canada}\; $^{2\;7}$\\
 \texttt{arna.ghosh@mail.mcgill.ca} \\
 \hspace{50mm}
   \And
 Zahraa Chorghay \footnotemark[1] \\
  \texttt{z.chorghay@gmail.com} \\
  \hspace{50mm}
  \And
 Shahab Bakhtiari $^{\ddag}$ \footnotemark[2]\; \thanks{Department of Psychology, Universit\'e de Montr\'eal, Montr\'eal, QC, Canada}\\
  \texttt{shahab.bakhtiari@umontreal.ca} \\
  \And
 Blake A. Richards $^{\ddag}$\; \footnotemark[2]\;\; $^2$\;\thanks{Department of Neurology \& Neurosurgery, McGill University, Montr\'eal, QC, Canada}\;\; \thanks{Montreal Neurological Institute, Montr\'eal, QC, Canada}\;\; \thanks{CIFAR Learning in Machines \& Brains Program, Toronto, ON, Canada}\;\; \thanks{\Google, Paradigms of Intelligence Team}\\
  \texttt{blake.richards@mcgill.ca} \\
}

\begin{document}
\maketitle
\begin{abstract}
Biological and artificial intelligence systems navigate the fundamental efficiency-robustness tradeoff for optimal encoding, i.e., they must efficiently encode numerous attributes of the input space while also being robust to noise. This challenge is particularly evident in hierarchical processing systems like the human brain. With a view towards understanding how systems navigate the efficiency-robustness tradeoff, we turned to a population geometry framework for analyzing representations in the human visual cortex alongside artificial neural networks (ANNs). In the ventral visual stream, we found general-purpose, scale-free representations characterized by a power law-decaying eigenspectrum in most areas. However, in certain higher-order visual areas did not have scale-free representations, indicating that scale-free geometry is not a universal property of the brain. In parallel, ANNs trained with a self-supervised learning objective also exhibited scale-free geometry, but not after fine-tuning on a specific task. Based on these empirical results and our analytical insights, we posit that a system’s representation geometry is not a universal property and instead depends upon the computational objective. 
\end{abstract}


\section{Introduction}
A fundamental challenge for both biological and artificial systems is navigating the tradeoff between robustness and efficiency. Efficiency, in the context of the ``efficient coding'' hypothesis in neuroscience, refers to reducing information redundancy by eliminating correlations in the system's input space \citep{barlow1961possible, atick1990towards, olshausen1996emergence, simoncelli2001natural}. This coding strategy results in a high-dimensional, sparse neural code, requiring only relatively simple downstream networks to read out complex features. Overall, an efficient neural code allows the system to capture numerous attributes of the input space and better utilizes its total representation capacity. On the other hand, a low-dimensional, correlated, and redundant coding strategy confers robustness in the presence of noise at the cost of reduced representation capacity \citep{shadlen1998variable, reich2001independent}. Since information processing requires representing increasingly complex features and abstractions of inputs --- often (if not always!) by navigating the efficiency-robustness tradeoff --- we turned to population geometry approaches in the human visual cortex and in artificial neural networks (ANNs) to understand the coding strategies used by different intelligent systems. 

\citet{stringer2019high} showed that the population geometry of neural responses in the mouse primary visual cortex (V1) has a unique signature: the eigenspectrum of the neural activity covariance obeyed a power law $n^{-\alpha}$, where $\alpha \sim 1$. These ``scale-free" representations are high-dimensional yet smooth, such that the coefficient of the power law, $\alpha$, reflects the fraction of neural variance that is devoted to representing coarse versus fine stimulus features. Examples of coarse distinctions are animate versus inanimate objects and outdoor versus indoor scenes, while fine distinctions could include differentiating individuals of the same species (e.g., faces of people, types of birds, etc.) or even different views of the same object. By having representations that are high-dimensional yet smooth, i.e., differentiable such that fine differences between stimuli can be represented while preserving coarse large-scale stimulus features, the neural code can navigate the tradeoff between efficiency and robustness. 

While scale-free representations have been observed in other species \citep{kong2022increasing, gauthaman2024universal} and in artificial neural networks (ANNs) \citep{agrawal2022alpha}, it remains an open question whether this geometric signature is a universal property or linked to a system’s specific computational objective. To address this question, we studied the representation geometry in both the human ventral visual cortex and ANNs, and present an analytical framework that links a system’s computational objective to its representation geometry. We find that representations in most visual areas are scale-free but in certain higher-order areas, representations lie in a subspace that does not have scale-free properties (i.e., a finite number of orthogonal directions are sufficient for capturing most of the stimulus-related information). This observation demonstrates that representations in the human brain are not universally scale-free.
In parallel, ANNs trained with self-supervised losses exhibited scale-free representations. When we finetuned these ANNs on a specific task, they exhibited representations without scale-free properties. Together, these observations lead us to postulate that the geometric shift that we observe in certain higher-order cortical areas and task-finetuned ANNs could be a hallmark of functional specialization, which is a key aspect of hierarchical information processing. Together, our results provide a parallel between biological and artificial systems, suggesting that the shift in representation geometry likely indicates a shift in the computational objectives in the respective information processing stages.

\section{Results}
\subsection{Representation geometry variation in the ventral visual cortical hierarchy}

\begin{figure}
    \centering
    \includegraphics[width=\linewidth]{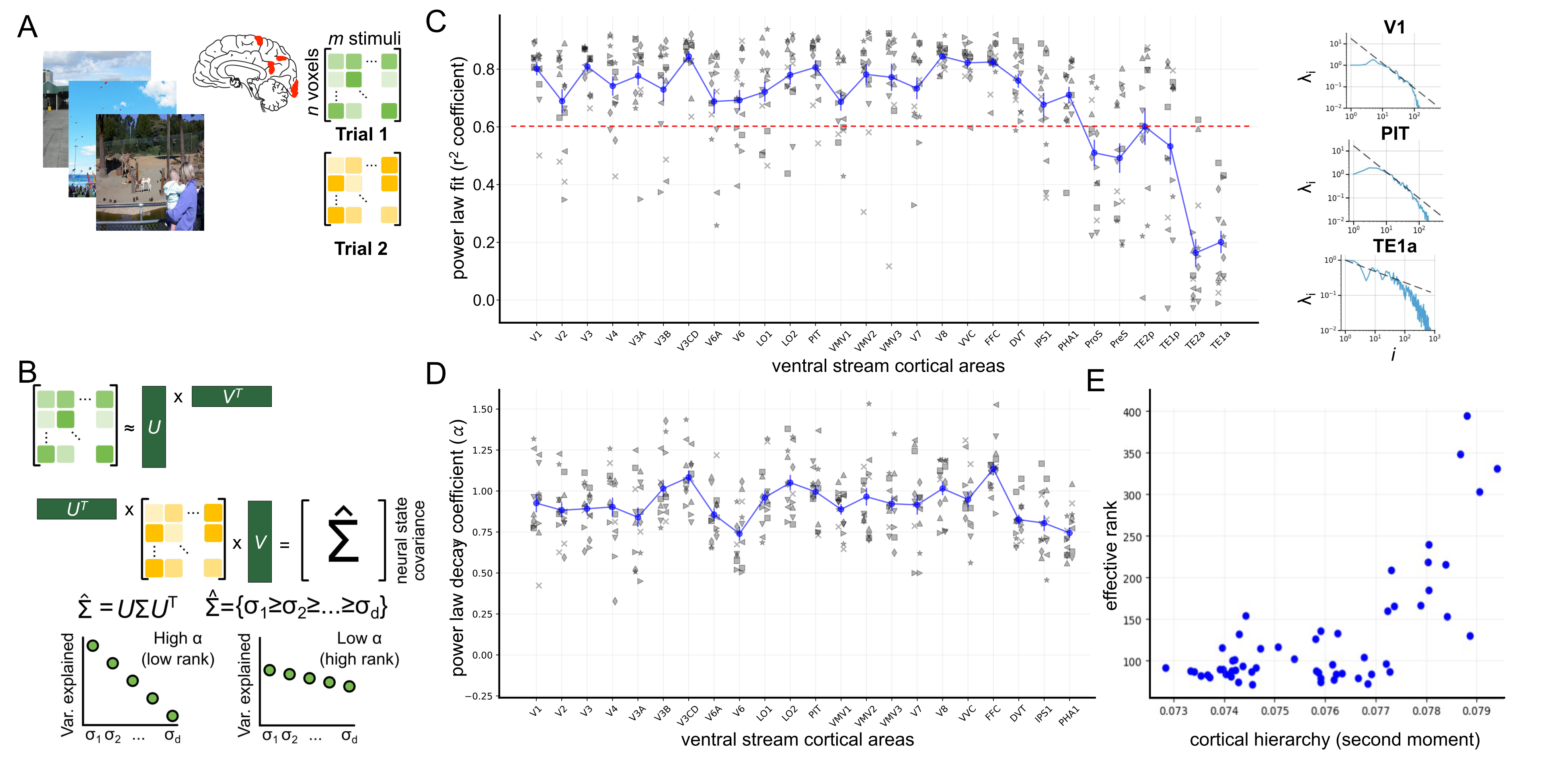}
    \caption{\textbf{Representation geometry across the human cortical hierarchy.} Early and intermediate areas, but not higher-order areas, exhibit scale-free representations. \textbf{{(A)}} Natural Scenes Dataset (NSD) and \textbf{(B)} cross-validation Principal Component Analysis (cvPCA) schematics. \textbf{(C)} Goodness of fit of the power law fit to the neural activity eigenspectrum for ventral visual cortical areas. \textbf{(D)} Decay coefficient ($\alpha$) for areas with a good power law fit ($r^2 > 0.60$). \textbf{(E)} Effective rank of neural activity in ventral visual cortical areas increases with increase in cortical hierarchy metric ($\rho=0.69$, $p<1e^{-6}$).(\textit{n} = left and right hemispheres for all 8 participants of NSD = 16 per area.)}
    \label{fig:fMRI_results}
\end{figure}

To investigate whether the human ventral visual cortex universally exhibits a power law code, we characterized the eigenspectrum of the activity covariance of individual brain areas using the Natural Scenes Dataset (NSD) (\Cref{fig:fMRI_results}a). Similar to \citet{stringer2019high}, we used cross-validation principal component analysis (cvPCA) to compute the eigenspectrum of the neural responses to natural images of the early and ventral visual stream cortex as defined by the HCP-MMP parcellation atlas (\citep{glasser2016multi})( \Cref{fig:fMRI_results}b). Next, we fitted a power law to the eigenspectrum decay, and characterized its fit using the $r^2$ coefficient (\Cref{fig:fMRI_results}c). Most cortical areas exhibited a power law fit ($r^2 > 0.6$), but some areas showed a poor power law fit ($r^2 < 0.6$), including the presubiculum (PreS), prostriate area (ProS), and the anterior and posterior areas of TE1 and TE2. Strikingly, in areas with a power law fit ($r^2 > 0.60$), $\alpha$ was close to 1 (\Cref{fig:fMRI_results}d) in line with existing literature \citep{stringer2019high, kong2022increasing, gauthaman2024universal}. In summary, while most areas of the ventral stream showed a scale-free geometry with $\alpha \sim 1$, certain higher-order areas did not, revealing that scale-free geometry is not a universal property across the brain. 

To further validate our findings, we turned to another useful metric for characterizing population geometry that has been used in the self-supervised learning (SSL) literature to quantify the expressivity of learned representations \citep{roy2007effective, garrido2023rankme}. This metric, called effective rank, is a distribution-agnostic metric of information geometry in high-dimensional spaces. We compared effective rank along the visual cortical hierarchy, with the hierarchy quantified by moment-based parameterization of staining intensity profiles ("second moment"; as per \cite{paquola2021bigbrainwarp}). We observed that the representations in higher-order cortical areas exhibit a higher effective rank. Along with \Cref{fig:fMRI_results}c, the effective rank indicates that higher-order cortical areas have high-dimensional representations that appear to lie in a subspace that does not have scale-free properties, i.e., most of the stimulus-related information is captured by a definite set of orthogonal directions.

\subsection{Variation in representation geometry may support task specialization}

Having observed differences in population geometry across the ventral visual hierarchy, we wondered what computational advantages this variation in representation geometry may confer? To address this question, we turned to artificial neural networks (ANNs). In contrast to the human brain, ANNs provide a more controllable experimental platform to examine the role of these two distinct population geometry coding strategies in supporting the computations performed by their respective systems.

\subsubsection{An analytical viewpoint}

Let us define a representation learning system that maps inputs from a domain $\mathcal{X} \subset \mathbb{R}^d$ to a learned feature space, $\mathbf{f} \subset \mathbb{R}^{d'}$. We adopt a theoretically ideal model to simplify the problem and postulate that each input $x \in \mathcal{X}$ is generated by a continuous and injective function $\mathbf{g}: \mathbb{S}^{D-1} \to \mathbb{R}^d$, where $\mathbb{S}^{D-1}$ is a sphere in $D$-dimensional latent space of latent factors $z$. These latent factors $z$ are assumed to encode the fundamental attributes, or a combination thereof, present in the stimulus. For instance, in the context of the natural image space $\mathcal{X}$, $z$ would encapsulate information regarding objects and their properties such as orientation or position. The function $\mathbf{g}(\cdot)$ then transforms these latent factors to synthesize the observed input $x$ that is subsequently processed by the representation learning system.

The goal of any representation learning system is to reliably recover (some or all) latent factors, $z$, from given input data $x$. When these latent factors of interest are known \textit{a priori}, the system aims to extract these specific factors while disregarding information about other factors. To the astute reader, this scenario may be reminiscent of the supervised learning setup, where the latent factors of interest are analogous to the predefined class identity of an image. Conversely, when latent factors of interest are not known \textit{a priori}, the representation learning system aims to identify as many latent factors as possible while also upholding specific symmetries. This situation is akin to the self-supervised learning (SSL) paradigm, in which symmetries are explicitly defined by incorporating specific invariance properties in the learned representation space, typically achieved through the application of data augmentations. These data augmentations are crucial for identifying which latent factors will be learned, as those that are preserved under the chosen set of data augmentations will be considered as task-relevant ``signal’’; augmentations that are not preserved will be considered ``noise,’’ and rejected or suppressed in the representation space. 

Despite these differences, both supervised and self-supervised learning leverage similar loss functions to train ANNs to learn effective representations. Supervised learning fundamentally relies on the cross-entropy loss to train ANNs, mapping input data to a probability distribution over predefined class identities. Recently, \citet{reizingercross} demonstrated a crucial theoretical insight: minimizing the cross-entropy loss was sufficient for learning representations that are a linear transformation of the underlying latent factor, $z$. This finding is particularly interesting if considering the connection to SSL, wherein there appears to be a fundamental equivalence between a diverse array of proposed loss functions \citep{zhaiunderstanding, ghosh2024harnessing} and the cross-entropy loss commonly used in contrastive SSL, such as SimCLR \citep{chen2020simple}. This SSL cross-entropy loss operates like an instance classification setup, where all symmetry-preserving augmented versions of a given image are required to be classified as belonging to the same instance. This SSL cross-entropy loss is also sufficient for learning a representation space that is a linear transformation of the latent factors preserved by the defined augmentations \citep{reizingercross}. Therefore, a unifying theoretical argument emerges: 
both supervised and self-supervised learning setups inherently promote the learning of representations that are linear transformations of the true factors. 
This fundamental link paves the way for a unified lens through which representations learned in both paradigms can be analyzed.
\begin{equation}
    \mathbf{f}(x) = \mathbf{f} \circ \mathbf{g}(z) = \mathcal{O}\Tilde{z}
\end{equation}
where $\mathcal{O}$ represents an orthogonal linear transformation, and $\Tilde{z}$ denotes the task-relevant subset of latent factors.

This theoretical finding relating the learned representation space to the latent factor space does not make specific assumptions about the ANN architecture or expressivity. Instead, it assumes that the network has sufficient capacity to learn a linear mapping and the training has converged to a minimum of the loss. When an eigenspectrum decomposition is applied to the learned representation space, it reveals the variance in the orthogonal directions within this space. Given that the learned representation space is a linear transformation of the task-relevant latent factor space, the representation eigenspectrum is indicative of the variance associated with the independent dimensions within the latent factors. Consequently, in a supervised learning setup, the representation eigenspectrum of ANNs will exhibit substantial variance corresponding to the latent factors essential for performing the task, followed by a sharp decline, signifying the suppression of other irrelevant latent factors. In contrast, the SSL representation eigenspectrum will demonstrate variance across all latent factors that the network successfully captured. The scale-free nature of this variance profile is very likely a consequence of the inherent structure of the latent factor space, a phenomenon further supported by recent studies on SSL learning dynamics \citep{simon2023more, ghosh2024harnessing}.
We will now empirically validate our analytical insights by studying the population geometry of representations learned in supervised and SSL setups.

\subsubsection{Empirical evidence}

To empirically test our core theoretical claim that the loss function dictates the geometry of the learned representation space, we compared ANNs trained with supervised and self-supervised learning (SSL) loss functions. We hypothesized that ANNs with general representations would exhibit a power law decay of the eigenspectrum, whereas specialized representations would not.

\begin{figure}
    \centering
    \includegraphics[width=\linewidth]{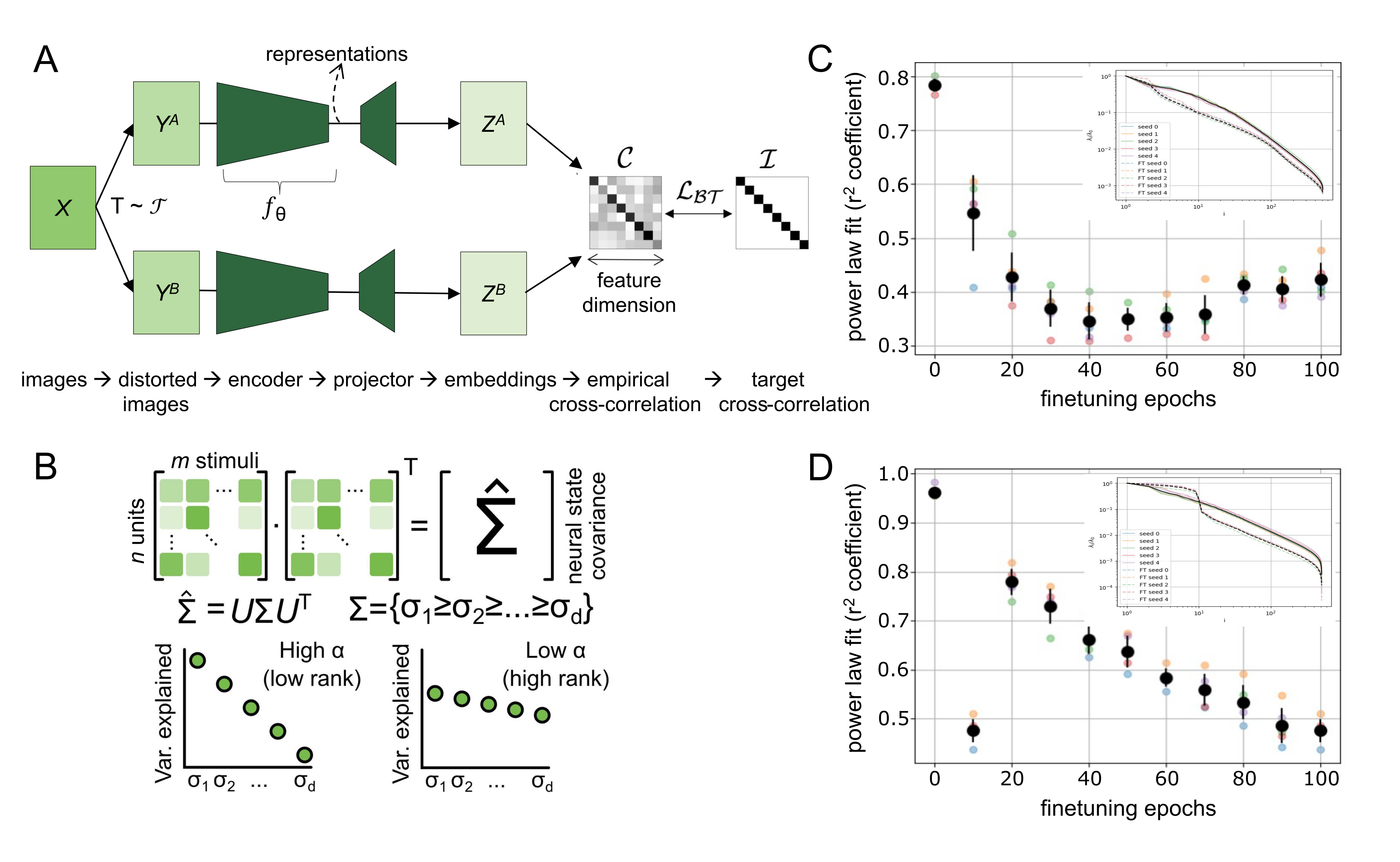}
    \caption{\textbf{Representation geometry changes as ANNs become specialized to a task.} General representations are scale-free but finetuned representations are not. \textbf{(A)} BarlowTwins training and \textbf{(B)} eigenspectrum computation schematics. Goodness of power law fit to the activity eigenspectrum across finetuning epochs on \textbf{(C)} ImageNet100 and \textbf{(D)} CIFAR-10.}
    \label{fig:SSL_results}
\end{figure}

Here, we pretrained a Resnet-18 network using the BarlowTwins objective \citep{zbontar2021barlow} on the Imagenet-100 dataset (\Cref{fig:SSL_results}a). As expected, these SSL-pretrained networks learned generic representations of naturalistic images and exhibited a strong power law decay in their representation covariance eigenspectrum (\Cref{fig:fMRI_results}b). To generate task-specialized versions of these networks, we finetuned the pretrained network using a cross-entropy loss to perform image recognition, resulting in an improved accuracy of $\geq 5\%$ on the downstream image recognition task (as compared to linear evaluation alone).  
While accuracy improved over finetuning epochs, these specialized networks no longer exhibited a good power law fit (\Cref{fig:fMRI_results}c). To further validate these findings, we also finetuned the BarlowTwins-pretrained networks to perform image recognition on a different dataset, CIFAR-10. We saw an impressive 20\% improved accuracy with finetuning, compared to linear evaluation alone. Following finetuning, these task-specialized networks also no longer demonstrated a good power law decay of their eigenspectrum (\Cref{fig:fMRI_results}d).

These findings demonstrate that when a network becomes specialized for a given task like image recognition, representations can shift away from being scale-free. This supports our theoretical claims about the link between a system’s representation geometry and the respective computation objective function.


\section{Discussion \& Conclusion}
In this work, we applied a population geometry framework to uncover fundamental computational principles of hierarchical information processing in both biological and artificial neural systems. We find that the representations are high-dimensional throughout the human ventral visual cortical hierarchy, as measured by the eigenspectrum decay and effective rank. Furthermore, we demonstrate a striking parallel between the ventral visual stream and artificial neural networks (ANNs): general-purpose representations, from most visual areas or from SSL pretrained networks, exhibit a scale-free geometry characterized by a power law decaying eigenspectrum. In contrast, specialized representations found in certain higher-order cortical areas and in ANNs finetuned to specific tasks lack this scale-free property. Together, our core finding is that the computational objective of a system is a key determinant of its representation geometry.

We posit that the scale-free geometry is a hallmark of robust, generic representation learning, analogous to SSL-pretrained ANNs. By capturing features at multiple scales, this coding strategy provides an inherent robustness to noise. It also allows representing fine stimulus details while preserving coarse, large-scale stimulus features \citep{stringer2019high, agrawal2022alpha}. We see such scale-free representations with $\alpha$ close to 1 in many ventral visual areas in the human brain \citep{stringer2019high, kong2022increasing, gauthaman2024universal} and in ANN models trained for generic representation learning \citep{agrawal2022alpha}. Overall, high-dimensional, scale-free representations are suited for supporting a wide variety of downstream behaviors.

On the other hand, there were some higher-order visual areas that did not show a power law decay of their eigenspectrum, demonstrating for the first time (to our knowledge) that scale-free representations are not universal in the human brain. 
This mirrors our ANN experiments, where finetuning a pretrained network to perform a specific image recognition task also eliminated the power law decay of the representation covariance eigenspectrum. Notably, such specialized representations (i.e., high-dimensional and not scale-free) have been shown to be computationally advantageous for tasks like few-shot learning \citep{sorscher2022neural}, where a system must quickly generalize to novel concepts with limited examples. 

These computational advantages conferred by the respective coding strategies suggest that they can be applied by intelligent systems for distinct objectives. This is particularly interesting in a neuroscience context, where our findings provide a computational perspective of how initial sensory areas ``solve'' for robust, generic representations, while higher-order unimodal and multimodal areas operate on this robust code to achieve increased functional specialization through the cortical hierarchy.

Our results provide a unified framework for understanding the link between a system's computational objective and its representation geometry. A critical next step will be to establish causal links between the computational objective and representation geometry, for instance, by investigating the specific training recipes and architectural constraints needed for the formation of a scale-free eigenspectrum. Furthermore, our findings suggest that the hierarchical structure of the brain may be an elegant solution to the dual challenge of achieving both generic, robust representations and efficient, task-specialized encoding. Our work hopes to inspire future research on a more mechanistic understanding of how hierarchical organization allows the brain to navigate the efficiency-robustness tradeoff, and how this biological solution might inform the design of better artificial intelligence systems.

\section{Methods}

\subsection{Neural dataset}

\paragraph{Experiment Design}To study the representation geometry across the visual hierarchy, we used the Natural Scenes Dataset (NSD). The NSD consists of high-resolution (7T) functional magnetic resonance imaging (fMRI) responses acquired from 8 participants performing a continuous recognition memory task on $\sim$$10,000$ natural scenes viewed over 30–40 scan sessions over one year \citep{allen2022massive}. The stimuli were obtained from the Microsoft COCO dataset \citep{lin2014microsoft}, containing complex everyday scenes with common objects. Images were shown three times over the course of the experiment, and trial-level response estimates were obtained for $\sim$$30,000$
stimulus presentations.

\paragraph{Preprocessing} We used the 1 mm volume preparation of the NSD and version 3 of the NSD single-trial betas (betas\_fithrf\_GLMdenoise\_RR). In this version, the haemodynamic response function was estimated for each voxel, the GLMdenoise technique was applied for denoising, and ridge regression was used to estimate the single-trial stimulus-evoked activity. The Human Connectome Project Multi-Modal Parcellation (HCP-MMP) atlas segment the brain into different cortical areas. Based on known literature, we identified the ventral visual areas and analyzed the data for these areas: visual areas 1-4 and 7 (V1, V2, V3, V4, V7); posterior inferotemporal complex (PIT); and occipital-temporal areas, including lateral temporal posterior areas 2 and 1 (TE2p, TE1p) and lateral temporal anterior areas 2 and 1 (TE2a, TE1a). For each area, similar to \citet{stringer2019high}, to preserve only task-relevant activity for each voxel, we performed spontaneous data cleaning by removing correlated activity across voxels that is shared during resting state (restingbetas\_fithrf) and task sessions.

\paragraph{Eigenspectrum Computation}
To estimate the eigenspectrum of neural activity, we performed cross-validation Principal Component Analysis (cvPCA) as per \citet{stringer2019high}. Briefly, we split the repeated responses for each unique stimuli into one of two groups, `Trial 1' and `Trial 2'. This split ensured that both groups had neural responses to the same stimuli set. Thereafter, we performed singular value decomposition of the `Trial 1' response matrix ($X_1$) yielding $X_1 = U\Sigma_1V^T$. Next, we computed the neural state covariance, $\Hat{\Sigma} = U^T X_2 V$, where $X_2$ denotes the `Trial 2' response matrix. The diagonal elements of $\Hat{\Sigma}$ were used as the neural covariance eigenspectrum. This process was repeated 5 times, each time with a different split of the repeats of stimuli presentation.

\paragraph{Cortical hierarchy metric} 
To quantify cortical hierarchy, for each area obtained from the BigBrain dataset \citep{amunts2013bigbrain} (\textit{n}=1 subject), we calculated the second moment of the cortical staining intensity profile as per \citet{paquola2021bigbrainwarp}. The cortical intensity profile indicates the relative density and depth of pyramidal neurons at each voxel, thereby providing an estimate of the cytoarchitecture-based laminar differentiation. A higher second moment indicates a higher-order cortical area.

\subsection{Spectral metrics}

\paragraph{Powerlaw decay coefficient, $\alpha$} To characterize the decay of the eigenspectrum, we tested whether it exhibits a heavy-tailed distribution, specifically, a power law decay, i.e., whether eigenvalues follow a distribution $\lambda_i\sim i^{-\alpha}$ \citep{stringer2019high, agrawal2022alpha}. We used a weighted linear regression in log space from rank 10 to $\sim$$100$, with weights as the inverse of log of the rank. We used the regression $r^2$ coefficient to estimate the goodness of power law fit. For areas with a good fit, we used the slope of the regression fit as the eigenspectrum's power law decay exponent, $\alpha$. 

\paragraph{Effective rank} The effective rank provides a quantitative measure of the effective number of dimensions that capture most of the variance. We measured the effective rank of the representation space as per \citet{roy2007effective}:
\begin{equation}
    \text{Effective\_Rank} := exp\left(-\sum_{k}{p}_{k}\log({p}_{k})\right) \quad \text{,} \quad p_k = \frac{\lambda_{k}}{\sum_{i}\lambda_{i}}+\in
\end{equation} 

\subsection{ANN experiments}

\paragraph{SSL pretraining} We trained a ResNet-18 on Imagenet-100 \citep{deng2009imagenet} with the BarlowTwins loss function \citep{zbontar2021barlow} for 100 epochs using the Adam optimizer \citep{kingma2014adam}, saving intermediate checkpoints at every 10 epochs. The pretraining loss function was as follows:
\begin{align}
    \mathcal{L}_{BT} &= \sum_i (C_{ii}-1)^2 + \beta \sum_i\sum_{j\neq i} C_{ij}^2 \nonumber\\
    C &= \frac{1}{n-1} \sum_{i=1}^{n} (f(\mathbf{x}_i) - \overline{f(\mathbf{x})})(f(\Tilde{\mathbf{x}}_i) - \overline{f(\Tilde{\mathbf{x}})})^T \nonumber \\
    \overline{f(\mathbf{x})} &= \frac{1}{n} \sum_{i=1}^{n} f(\mathbf{x}_i) \quad \text{,} \quad \overline{f(\Tilde{\mathbf{x}})} = \frac{1}{n} \sum_{i=1}^{n} f(\Tilde{\mathbf{x}}_i) 
    \label{eq:background_BT}
\end{align}
For each model checkpoint, we computed $\alpha$ and effective rank of the final layer representations. We found that the eigenspectrum of representation covariance matrix at each checkpoint had a good power law fit ($r^2 > 0.9$).

\paragraph{Linear evaluation} We appended a linear layer that mapped the learned features to class logits and trained this linear layer on the classification loss. The parameters of the rest of the network were frozen, thereby keeping the representation space unchanged. For the Imagenet-100 dataset, the output of the linear layer was 100-dimensional, whereas it was 10-dimensional for CIFAR-10 finetuning. In both cases, the linear layer was trained for 200 epochs with Adam optimizer.

\paragraph{Supervised finetuning} We appended a linear layer that mapped the learned features to class logits and finetuned the BarlowTwins-pretrained network along with the linear layer on the classification loss. As with linear evaluation, the output of the linear layer was 100-dimensional for the Imagenet-100 dataset, whereas it was 10-dimensional for CIFAR-10 finetuning. In both cases, we finetuned the network for 200 epochs with Adam optimizer.

\paragraph{Eigenspectrum computation} In each case, the representations were extracted from the final layer of ResNet-18 --- before the projector network (used during BarlowTwins pretraining) or the linear classification layer (used during finetuning). We computed the representation covariance matrix for 10,000 stimuli and performed eigenspectrum decomposition of this covariance matrix. As for the neural covariance eigenspectrum, the spectral metrics were computed using this representation covariance eigenspectrum.

\section*{Acknowledgements}
We would like to thank Sethu Kovendhan Boopathy Jegathamabal for facilitating access to the BigBrain dataset, and Casey Paquola for her insights regarding cortical hierarchy metrics.

A.G. was supported by Vanier Canada Graduate Scholarship. S.B. was supported by NSERC (Discovery Grant RGPIN-2023-03875) and Canada Excellence Research Chair (CERC). B.A.R. was supported by NSERC (Discovery Grant RGPIN-2020-05105; Discovery Accelerator Supplement RGPAS-2020-00031) and CIFAR (Canada AI Chair; Learning in Machines and Brains Fellowship). 
We also acknowledge the material support of NVIDIA
in the form of computational resources, as well as the compute resources, software, and technical help provided by Mila (mila.quebec).

\bibliographystyle{plainnat}
\bibliography{references}

\clearpage
\appendix
\section{Additional spectral metric results}
\label{apd:first}



\begin{figure}[!htbp]
    \centering
    \includegraphics[width=\linewidth]{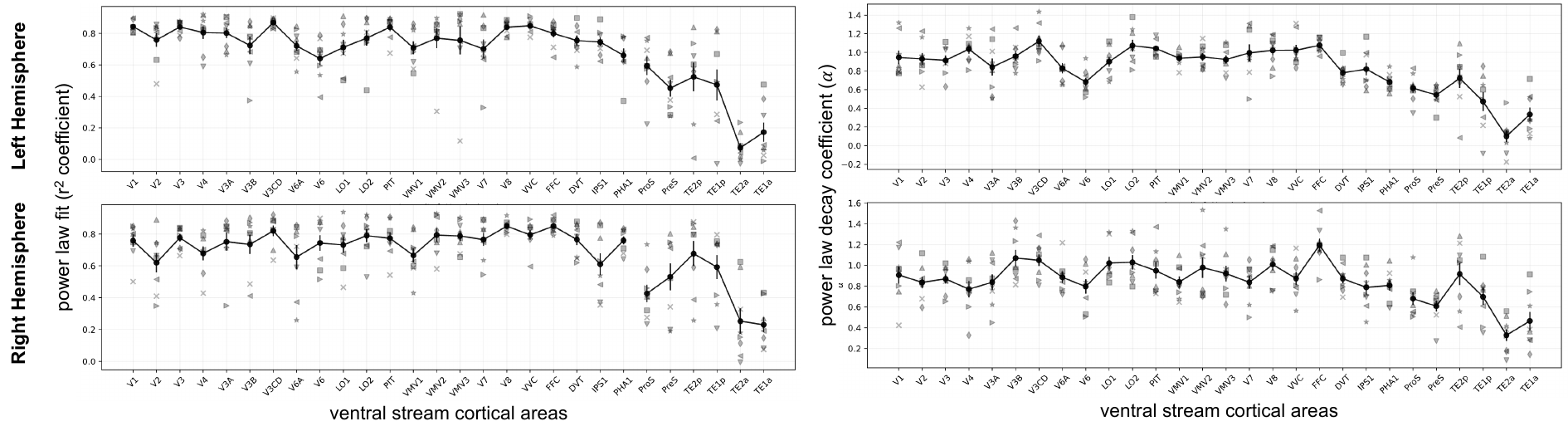}
    \caption{Power law goodness of fit, measured using regression $r^2$ coefficient, and decay coefficient ($\alpha$) for different areas for each hemisphere.}
    \label{fig:powerlaw_hemispheres}
\end{figure}

\begin{figure}[!htbp]
    \centering
    \includegraphics[width=\linewidth]{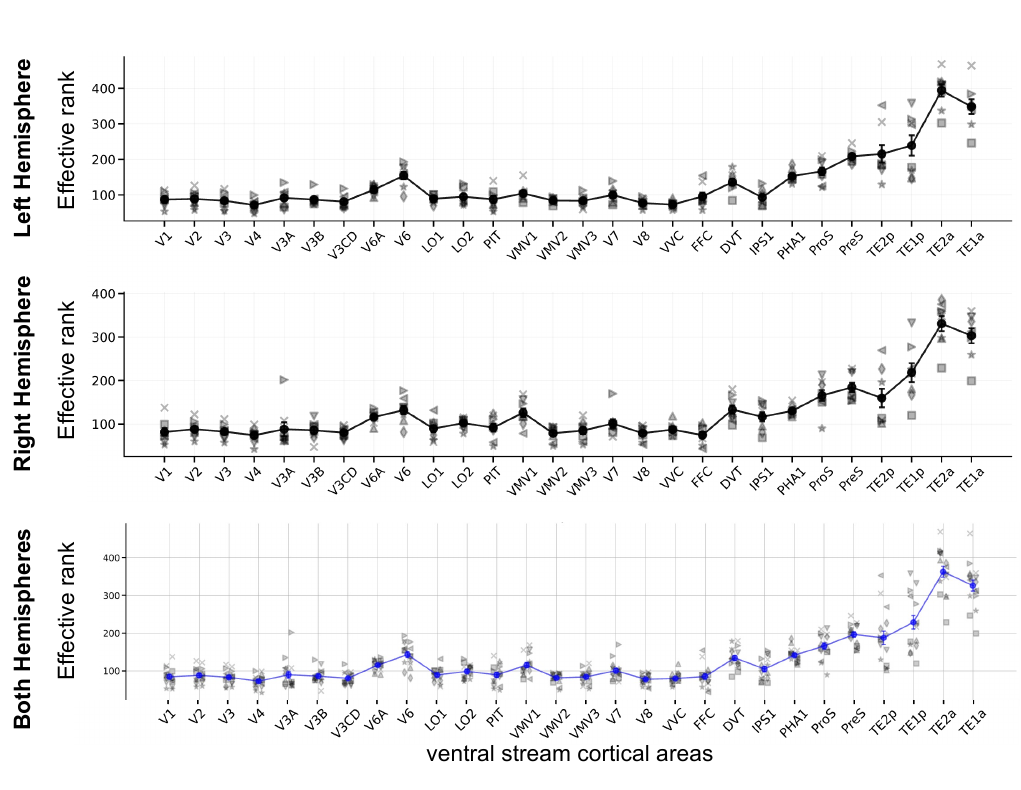}
    \caption{Effective rank for different areas for each hemisphere, and combined for both hemispheres.}
    \label{fig:rankme_hemispheres}
\end{figure}

\end{document}